\documentstyle[12pt]{article}
\begin{document}
\textwidth 16.cm 
\textheight 21.cm 
\pagestyle{empty}
\renewcommand{\baselinestretch}{1.2}                                             
\newcommand{\be}{\begin{equation}}                                              
\newcommand{\ee}{\end{equation}}                                                
\newcommand{\bc}{\begin{center}}                                                
\newcommand{\ec}{\end{center}}                                                  
\newcommand{\vs}{\vspace{1.0 cm}}
\newcommand{\vm}{\vspace{.5 cm}}
\newcommand{\hs}{\hspace{1.0 cm}}
\begin{titlepage}                                                               
\bc

{\Huge A STUDY OF NANOCRYSTALLINE COBALT PREPARED BY BALL MILLING \\ }
                                                                                
\vs
\vm
{\large  J.C. de Lima, V. H. F. dos Santos, T. A. Grandi and R.S. de Biasi$^{a}$}
 \vm

{\em 
Departamento de F\'{\i}sica, \\ 
Universidade Federal de Santa Catarina, \\
UFSC, Florian\'{o}polis CEP 88040-900, SC, Brazil. }

{\em 
$^a$ Departamento de Engenharia Mec\^anica e de Materiais, Instituto Militar de 
Engenharia, Rio de Janeiro, Brazil \\ }

\ec
\end{titlepage}                                                                 

\newpage                                                                        

\bc
{\Large {\bf Abstract}}
\ec

\vspace{1.0 cm}

The hcp-fcc transformation induced in cobalt powder by ball milling at room 
temperature was studied using the excess Gibbs free energy for the metals, in
nanometric form, to calculate the activation barriers that the atoms must 
overcome for the phase transformation to occur. The crystalline component of 
the nanocrystaline cobalt produced in our laboratory, via ball milling, was
analized using the radial distribution function technique. The results suggest
that the crystalline component has the same structure as the bulk, contradicting
results already reported.

\newpage
\bc
{\bf 1. Introduction}
\ec
\vs

The advent of nanostructured materials has made it possible to imagine a wide
range of new technologies. This is associated with the presence of atomic
arrangements located at defect centers, such as grain boundaries and interphase
boundaries, allowing one to produce solids with new atomic structures and 
physical properties.

From the structural point of view, nanostructured materials can be regarded as 
being made up of two components, one crystalline, with dimensions on the order
of some nanometers, that preserves the structure of bulk crystal, and another
interfacial, composed of defects (grain boundaries, interphase boundaries,
dislocations, etc.). The latter can be regarded as a highly disordered phase,
lacking the short-range order present in glasses as well as the long-range order
found in the crystalline state$[1,2]$. This component has caused controversy in
the literature. Gleiter${'}$ analysis$[3]$ is based on a gaseous model, while
other authors$[4]$ think this is not the best approach. The number of atoms in
the two components is about the same$[5]$.

Recently, Huang et al.$[6]$ reported a study of phase transformation induced in
cobalt powder by ball milling at room temperature. The main conclusions reached
by the authors are: for a milling intensity below a lowe threshold value, the
starting mixture of hcp and fcc phases transforms to a single hcp phase; for an
intermediate range of milling intensities there is a second transformation of the
hcp phase to a mixture of the fcc and hcp phases; finally, for milling intensity
above an upper threshold value, there is a third transformation of the fcc and
hcp mixture to a single fcc phase. Previously, a detailed thermal and structural
study of the same phase changes had been performed by Mazzone$[7,8]$; the results
are not in agreement with those of Ref.$[6]$. According to Mazzone, the structural
transformations induced by ball milling of cobalt powder are more subtle because
of two factors: $(1)$ the progressive formation of an increasingly more faulted
hcp phase which eventually transforms into a structure based on an almost random
stacking of close-packed planes; $(2)$ the progressive contamination from the
steel milling tools, which for an iron concentration of the order of a few atoms
per cent stabilizes the fcc phase, as shown by the Fe-Co phase diagram.

There is in the literature a study performed by Babanov et al.$[9]$ applying
the EXAFS technique to nanocrystalline cobalt. These investigators measured the
EXAFS oscillations of polycrystalline, nanocrystalline and grain boundary
components of cobalt powder. Using the regularization method, they found the
corresponding pair correlation functions. For nanometric cobalt, they reported
a strong reduction in the nearest-neighbor coordination number (6.35 atoms)
of the crystalline component in comparison with polycrystalline cobalt (13.1
atoms) while 3.65 atoms located at 0.252 nm were found for the grain boundary
component.

According to the basic idea that the crystalline component of nanocrystalline
materials preserves the structure of the bulk crystal, a strong reduction in
the nearest-neighbor coordination number of this component, as the one observed
by Babanov, is not expected, since it would imply that its structure is different
from that of the bulk crystal. A recent study performed by us $[5]$ on nanocrystalline
nickel, using the radial distribution function (RDF), shows that the nearest-
neighbor coordination number of this component is similar to that of the bulk
crystal.

In this paper, we investigate the hcp-fcc transformation in nanocrystalline
cobalt obtained via ball milling in our laboratory, using the excess Gibbs
free energy for the nanometric cobalt and RDF analysis to obtain the nearest-
neighbor coordination number of the crystalline component.

\vs
\bc
{\bf 2. Experimental Procedure}
\ec
\vs

Cobalt (purity, $99.5\%$) was milled for 62 h in argon atmosphere in a Spex
8000 shaker mill, using spheres and a cylindrical steel vial. The mass ratio
between the ball and the powder mass (BPR) was 4:1. A ventilation system was used
to keep the temperature of the vial close to room temperature. The milled 
powder samples were analyzed by X-ray diffraction using a Siemens difractometer
working with the Cu K$\alpha$ line ($\lambda = 0.15418 nm$).

\newpage
\bc
{\bf 3. Results and Discussion}
\ec
\vs

Fig. 1 shows the X-ray diffraction patterns (XRD) for the cobalt as-received
(spectrum A) and after 62 h of milling (spectrum B). From spectrum A, using 
the integrated intensities of the peaks, we obtained $64.9\%$ and $35.1\%$
of hcp and fcc phases, respectively. Spectrum B still shows both phases, but the
fraction of the fcc phase is larger than in spectrum A. This result agrees
quite well with those of Huang et al.$[6]$ for an intermediate range of milling
intensities. Spectrum B does not show other peaks besides those associated
with the fcc and hcp phases, indicating that if there is contamination from
the steel tools, it must be very low. In recent study $[5]$, using a Spex 8000
shaker mill, the same number of steel balls and the same BPR, we measured by
X-ray energy dispersive analysis a contamination of less than $3\%$ for nickel
powders after 30 h of milling.

In order to explain the hcp-fcc transformation of nanocrystalline cobalt upon
milling, we will use the thermodynamic properties and the stability of grain
boundaries in nanometric metals and the concept of lattice stability.

Ball milling has been used as an alternative technique to obtain some metals,
such as Ni, Fe, Pd, Co and Cu, in nanometric form. In the first stage of milling,
there is a drastic reduction in crystallite size. Further milling leads to ultrafine
crystallites. X-ray analysis shows that these crystallites have a few nanometers
in diameter. The reduction of crystallite size to a few nanometers cannot be
accomplished without creating an interfacial component.

From the thermodynamic point of view, a nanometric powder is in a metastable
state whose Gibbs free energy is greater than that of the crystalline state. Without
annealing, there is no spontaneous transition to the crystalline state, suggesting
the existence of an activation energy .

Using an universal equation of state proposed by Rose et al.$[10]$ to study the
thermodynamic properties and the stability of grain boundaries in nanometric
metals, Fecht showed$[11]$ that the excess Gibbs free energy of the interfacial
component grows with the excess volume but has a local minimum due to the maximum
entropy that this component can reach. The entropy of sublimation of a gas was
taken as this limit. The local minimum is responsible for the stability of the metals
in nanometric form. From Fechet${'}$ paper is clear that for a nanocrystalline-
to-crystalline trasition to occur, the atoms located at the interfacial component
must overcome an activation energy  whose value can be calculated from the values of 
the maximum and minimum excess Gibbs free energies. Thus, in order to promote
the transition, it is necessary to introduce a certain minimum amount of external
energy; this can be done by annealing the material at an appropriate temperature.
As an illustration, we show in Fig. 2 the excess Gibbs free energy, given by equation
$\Delta$G(V,T) = $\Delta$H(V,T) - T$\Delta$S(V,T), for nanocrystalline Co (hcp)
at 300 K. The $\Delta$V = 0.0 value corresponds to the crystalline state. As
can be seen, there is a local minimum at $\Delta$V = 0.410; the value of the
activation energy  is 1.05 x 10$^{-20}$ J/at or about 0.066 eV/at.

Unfortunately, we did not find in the literature enough data to calculate the
excess Gibbs free energy and the excess volume for nanocrystalline Co in the
fcc phase. However, we can use the concept of lattice stability to estimate the
Gibbs free energy of this phase. The lattice stability parameter of an element
is taken as the difference between the free energy of the phase that is being
described and the free energy of the other phase, used as a reference$[12,13]$.
The lattice stability parameter for cobalt (hcp) with respect to cobalt (fcc) is

	Co$^{fcc \rightarrow hcp}$ = [-460 + 0.628 T]  (J mol$^{-1}$)

At 300 K, the value of the lattice stability parameter is about -271.6 J mol$^{-1}$
or -28 x 10$^{-4}$ eV/atom, i.e., a very small value. As the temperature increses,
the lattice stability parameter becomes even smaller. The Gibbs free energies of the
hcp and fcc phases are thus almost equal. On the other hand, the results shown
in Fig. 1 seem to suggest that nanocrystalline Co in the fcc phase has an activation energy
 greater than that for nanocrystalline Co in the hcp phase, since the fcc-hcp
transformation is not observed.

In order to transform to the fcc phase, the Co atoms belonging to the interfacial
component of the hcp phase have to overcome an activation energy ; the required
energy is provided by collisions with the balls. In a Fritsch Pulverisette 5
planetary ball mill, the local temperature can reach up to 680 K$[14]$ and
several alloys have been formed using this type of mill.In a Spex mill, the
local temperature can be even larger. Assuming that T = 680 K in the expression
$E_{b}$ =$K_\beta$T = 8.617 x 10$^{-5}$ T eV/K, we obtain an average energy of
about 0.059 eV/atom, which agrees quite well with the activation energy  calculated
at 300 K for nanometric cobalt (hcp). However, based on recent studies$[15]$,
we believe that the value of the activation energy  depends on the average temperature
at the time when the nanometric powder was formed. We also believe that other
processes may contribute to the energy balance, as the following arguments
suggest: (1) During ball milling of nickel, even after the crystallite size has
stabilized at about 15 nm for 30 h of milling, the crystalline component continues
to store energy, reaching a saturation value of about 120 J g$^{-1}$ (about
0.073 eV/at) after 120 h of milling$[5]$. The same must occur during ball milling
of cobalt. (2) If part of the energy comming from the balls is used for the
annealing of defects in both hcp and fcc crystalline phases, the energy thus
released is available for the transformation process; in the next collision, new defects
are created, generating a continuous cycle. Thus, these two processes (collisions
with the balls and annealing of the defects) are the necessary ingredients for
the energy balance which defines the driving force of hcp-fcc transformation of 
cobalt upon milling. Of course, when the milling time is increased, the amount
of the hcp phase decreases, reducing the energy associated with the annealing
ofdefects of this phase. This reduction could explain the presence of a mixture
of the fcc and hcp phases after 62 h of milling (see Figure 1, spectrum B) as
observed by Huang et al.$[6]$. Thus, since the lattice stability parameter
calculated above is negligible, the fcc phase will be stable.

The influence of mechanical milling on the formation of nanocrystalline cobalt
was investigated using radial distribution function (RDF) analysis$[16]$. The
structure factors, I(K), for as-received and milled powders as a function of
the wave vector K = 4$\pi$ sin$\theta/\lambda$, were obtained from XRD data
using the method described in Ref.$[17]$; the results are shown in Fig. 3. The
anomalous dispersion terms f${'}$ and f${''}$ for Co atoms$[18]$ were taken into
account. A comparison between the I(K) functions for a sample milled for 62 h
and for as-received cobalt powder shows a reduction in the highest peak
intensity of about $49\%$. This reduction is attributed to the interfacial
component (containing about $49\%$ of the total number of atoms) whose contribution
to the XRD pattern is diffuse. Fig. 4 shows the radial distribution functions
(RDF) obtained from the Fourier transform of the function F(K)=K[I(K)-1]
multiplied by a damping factor of the form exp(-$\alpha^{2}$K$^{2}$), with
$\alpha^{2}$ = 0.01 nm$^{2}$, in order to impose the convergence of F(K). On
the RDF curve, the average interatomic distance and the coordination number for
each atomic shell are given by the position of the maximum and the area under 
the peak, respectively. In spite of termination effects due to the limited range
of the data (K$_{max}$ = 0.6225 and 0.7050 nm$^{-1}$ for the structure factor
corresponding to as-received and milled cobalt powder, respectively), we observe
that the RDF${'}$s are of good quality. For as-received cobalt, the RDF (solid line)
shows that the first neighbors shell is asymmetrical and not well isolated. 
A fit (dotted line) can be made assuming two Gaussian distribution functions of
12 Co neighbors centered at 0.248 nm and 3 Co neighbors centered at 0.326 nm.
In order to do this, we explicitly take into account the K dependence by
Fourier transforming into the K space the Gaussian distributions of distances
and back Fourier transforming over the same K range as the one used to calculate
the RDF${'}$s. Using the same procedure, the RDF (crossed line) obtained for
cobalt milled for 62 h was fitted assuming one Gaussian distribution function
of 12.9 Co neighbors centered at 0.258 nm. The slight shift of the interatomic
distance to larger values and the increase of the coordination number after
milling are probably due to strain fields extending from the boundary regions
to the crystallites, displacing the atoms from their ideal lattice sites$[15]$.

Our results for nanocrystalline cobalt do not agree with those of Babanov et
al.$[9]$, who found a nearest-neighbor coordination number of 6.35 atoms for
the crystalline component of nanocrystalline cobalt. The results obtained in
this study confirm that the crystalline component of nanocrystalline materials
preserves the structure of bulk crystal, as expected.

\vs
\bc
{\bf Acknowledgements}
\ec
\vs

We thank the Conselho Nacional de Desenvolvimento Cient\'{\i}fico e 
Tecnol\'ogico (CNPq) and the Ag\^encia Financiadora de Projetos (FINEP)
, Brazil, for financial support. 

\newpage
\bc
{\bf References}
\ec
\vs

[1] X. Zhu, R. Birringer, U. Herr and H. Gleiter, Phys. Rev. B, 35 (1987) 9085.

[2]  H.J. Fecht, Acta Metall., 38 (1990) 1927.

[3]  H. Gleiter, NanoStruct. Mater., 1 (1992) 1. 

[4]  E.A. Stern, R.W. Siegel, M. Newville, P.G. Sanders and D. Haskel,
Phys.Rev. Lett., 75 (1995) 3874.

[5]  T.A. Grandi, V.H.F. dos Santos and J.C. De Lima, Solid State Commun.,
112 (1999) 359. 

[6]  J.Y. Huang, Y.K. Wu and H.Q. Ye, Appl. Phys. Lett., 66 (1995) 308. 

[7]  F. Cardellini and G. Mazzone, Philos. Mag. A67 (1993) 1289. 

[8]  G. Mazzone, Appl. Phys. Lett., 67 (1995) 1944. 

[9] Yu. A. Babanov, I.V. Golovshchikova, F. Boscherini, T. Haubold and S.
Mobilio, Physica B 208-209 (1995) 140. 

[10] J.H. Rose, J.R. Smith, F. Guinea and J. Ferrante, Phys. Rev. B 29 (1984) 2963.

[11] J.H. Fecht, Acta Metall. Mater., 38 (1990) 1927. 
 
[12] L. Kaufman and H. Bernstein, Computer Calculation of Phase Diagrams,
Academic Press, New York and London, (1970). 

[13] N. Saunders and A.P. Miodownik, J. Mater. Res., Vol. 1, No. 1 (1986) 38.

[14] L. Schultz, J. Less-Common Met., 145 (1988) 233. 

[15] J.C. De Lima, T.A. Grandi, V.H.F. dos Santos, P.C.T. D${'}$Ajello and
A. Dmitriev, Phys. Rev. B 62 (2000) 8871. 

[16] Y. Waseda, Novel Application of Anomalous (Resonance) X-ray Scattering
for Structural Characterization of Disordered Materials, Springer-Verlag, Berlin,
edited by H. Araki, J. Ehlers, K. Hepp, R. Kippenhaln, H.A. Weindenmuller and
J. Zittariz, (1984). 

[17] C.N.J. Wagner, Liquid Metals, edited by S.Z. Beer, Marcell Dekker, Inc.,
New York, (1972) pp. 258-329. 

[18] S. Sasaki, Anomalous Scattering Factors for Synchrotron Radiation Users,
Calculated using Cromer and Liberman${'}$s Method, National Laboratory for
High Energy Physics, Japan, (1984). 

\newpage 

\vspace{1.0 cm}

\bc
{\bf Figure Captions}
\ec

\vspace{1.0 cm}

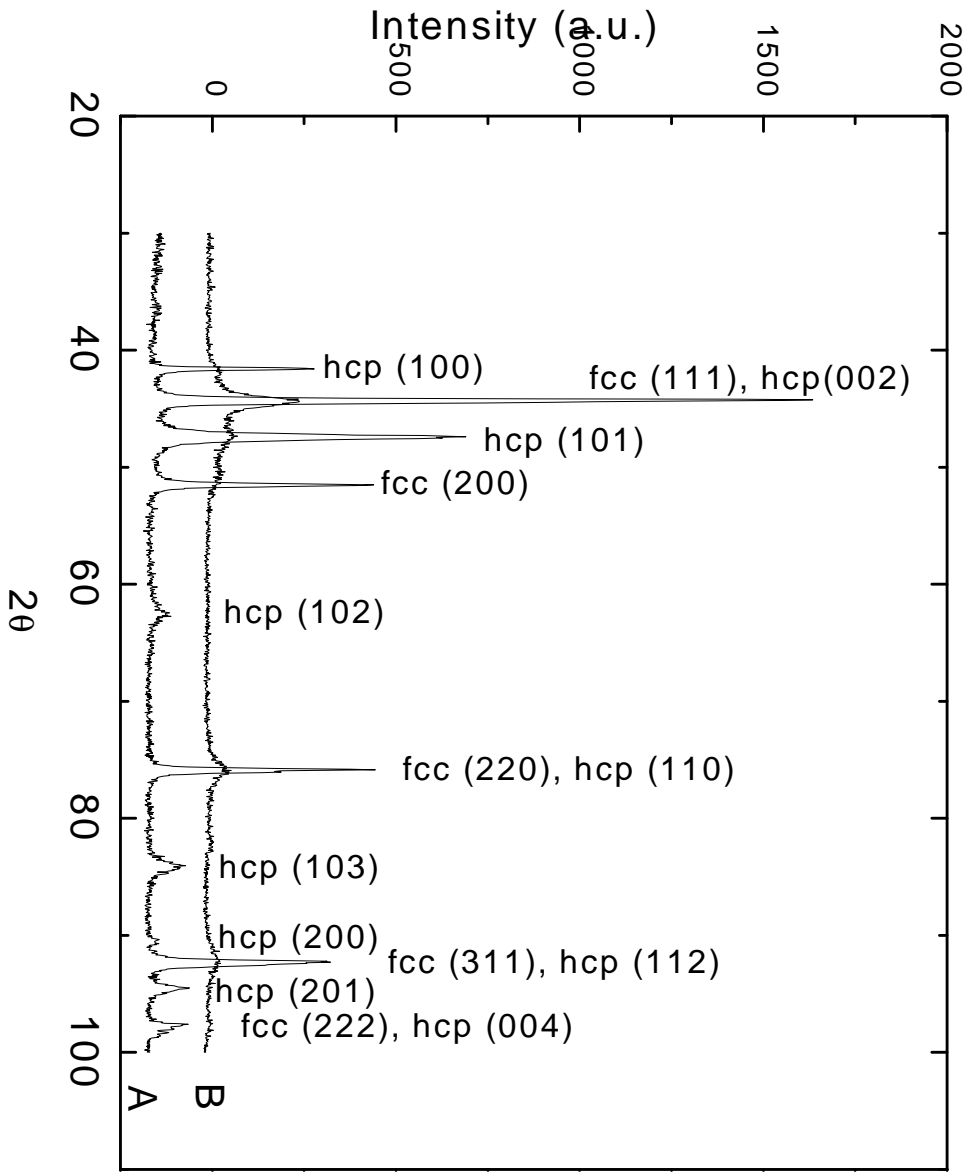
Figure 1 : Experimental X-ray diffraction patterns of the cobalt powders:
before milling (A) and after milling for 62 h (B). 

\vspace{1.0 cm}

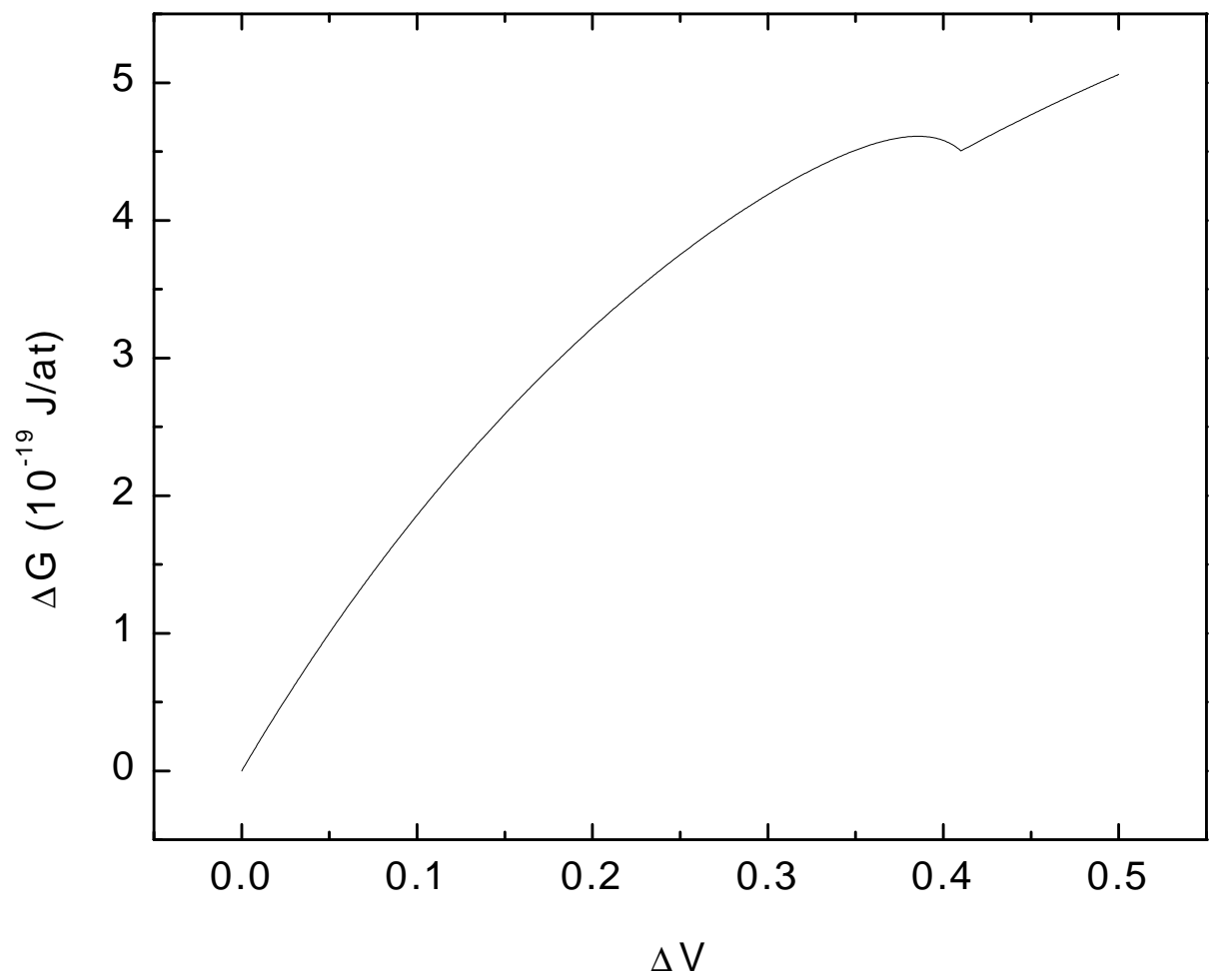
Figure 2 : Excess Gibbs free energy as a function of excess volume for 
nanocrystalline Co (hcp) at 300 K. 

\vspace{1.0cm}
Figure 3 : Structure factors of cobalt powders: before milling (dotted line) and
after milling for 62 h (solid line). 

\vspace{1.0 cm}
Figure 4 : Radial distribution functions of cobalt powders: before milling 
(solid line) together with a fitting (dotted line) assuming two Gaussian
distributions (see text) and after milling for 62 h (crossed line).

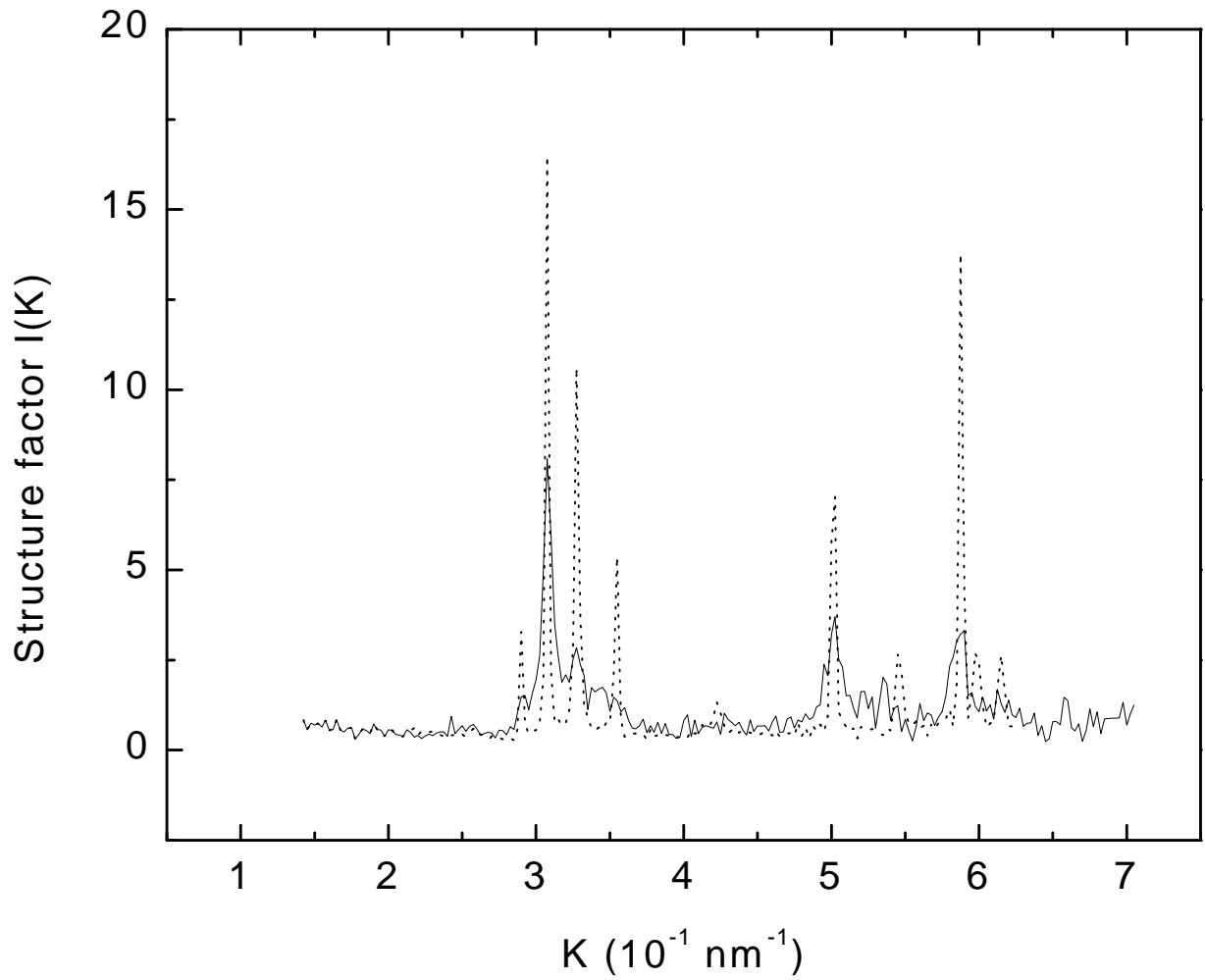
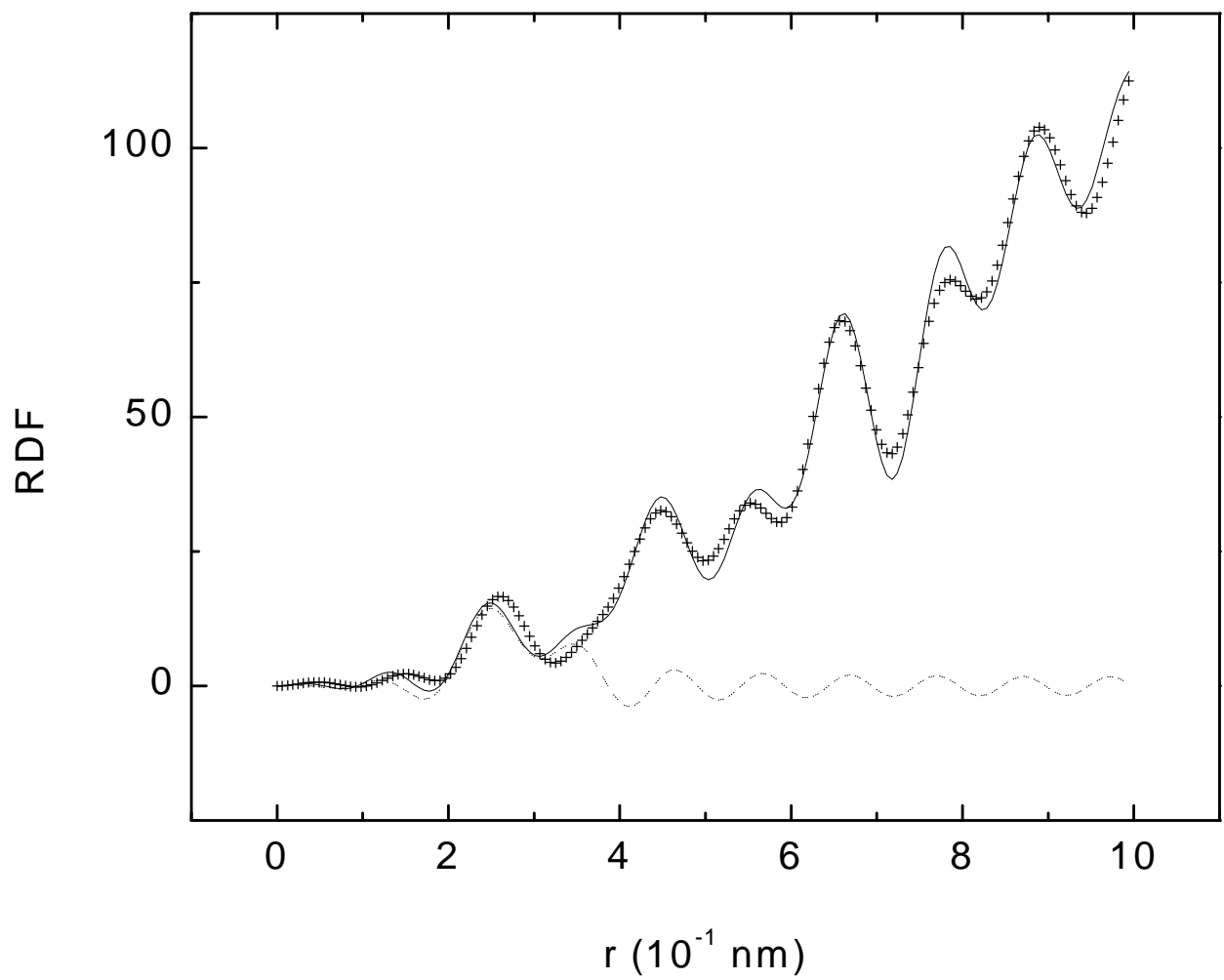
\end{document}